\documentclass[12pt,noshowpacs,nofootinbib,notitlepage,amsmath,amssymb]{revtex4-1}

\usepackage[utf8]{inputenc}
\usepackage[english]{babel}
\usepackage{tocvsec2}
\usepackage{amsthm} 
\pdfoutput=1
\allowdisplaybreaks
\usepackage{graphicx,color}
\usepackage[colorlinks=true,citecolor=blue,linkcolor=blue,urlcolor=blue]{hyperref}
\usepackage{comment}
\usepackage{appendix}
\usepackage{caption}
\usepackage{subcaption}
\usepackage{footmisc}
\usepackage{bm}
\usepackage{dcolumn}
\usepackage{gensymb}
 \usepackage{float}
\restylefloat{table}
\usepackage{multirow}
 \usepackage{hyperref}
\hypersetup{
    colorlinks,
    citecolor=black,
    filecolor=black,
    linkcolor=blue,
    urlcolor=blue
}
\usepackage{lipsum}
\usepackage{titlesec}
\usepackage{slashed}
\newcommand\scalemath[2]{\scalebox{#1}{\mbox{\ensuremath{\displaystyle #2}}}}
 
\newcommand{\Tr}{\ensuremath{\operatorname{Tr}}}
\numberwithin{equation}{section}

\makeatletter
\renewcommand\@seccntformat[1]{\csname the#1\endcsname.\quad}
\makeatother

\makeatletter
\renewcommand\p@subsubsection{} 
\makeatother
\newcommand*{\TitleFont}{%
      \fontsize{19}{20}%
      \selectfont}
\usepackage{setspace}

\begin{document}

\title{\TitleFont Instability of Chern-Simons Theory with Fermions\\ at Large N\vspace{1.0cm}}

\vskip 1cm
\author{ \large Chen Zhang\vspace{0.2 cm}}

\email{czhang@physics.utoronto.ca}

\affiliation{Department of Physics, University of Toronto, Toronto, Ontario, Canada M5S 1A7\vspace{0.5cm}}

\begin{abstract}
We study the (in)stability around the dynamical gap solution of the $U(N)$ Chern-Simons gauge theory with fundamental fermions (massless or massive) coupled in $D=3$ at large $N$. Explicit analyses on both the Auxiliary-Field (AF) and the Cornwall-Jackiw-Tomboulis (CJT) effective potentials are given. In both approaches we manage to analytically identify the saddle-point instability around the gap solution. We also give a comparison with the QCD-like theories. This study can help understanding the scale symmetry breaking picture of this theory.
\end{abstract}

\maketitle
\clearpage

\begin{spacing}{1.7}
\tableofcontents
\end{spacing}
\clearpage
\section{Introduction}
\label{intro}

The $D=3$ Chern-Simons gauge theory has a direct analog with $D=4$ QCD. Both are classically scale invariant and have great simplification at large $N$, but $D=3$ Chern Simons theory can keep its conformality and can be exactly solved in leading order of the large $N$ expansion. Related studies with matter in the fundamental representation coupled in this theory have made a lot of progress~\cite{Giombi}$\sim$\cite{CS_Thermal}. 

In recent works on this theory with massive fundamental fermion phase, there are some debates on whether there exists a scalar bound state like dilaton. Bardeen~\cite{Bardeen} argued that a dilaton, which is the pseudo-Goldstone boson associated with the spontaneous breaking of scale invariance, exists. However Ref.~\cite{Jain, Frishman2, Moshe} argued there is no such bound state appear in the spectrum of this theory. This controversy calls for a further understanding of the symmetry breaking picture of this theory. Motivated by this, we shall study the stability of this theory's effective potential around its gap solution (dynamical vacuum). The fermions can be in massive phase, where the scale symmetry is dynamically broken, or massless phase, where the scale symmetry is retained. We'll first study massive fermion phase, and later argue that the massless phase does not alter our conclusion of instability. 

In this section a brief introduction is given, mainly referring to Ref.~\cite{Giombi, Bardeen, Frishman2}.
The Euclidean action in this theory is given by
\begin{equation}
S = \frac{{i\kappa }}{{8\pi }}{\varepsilon ^{\mu \nu \rho }}\int {{d^3}x\,Tr\left( {A_\mu ^a{\partial _\upsilon }A_\rho ^a + i\frac{2}{3}A_\mu ^aA_\upsilon ^bA_\rho ^c{f^{abc}}} \right) + } \int {{d^3}x} \left[ {{\psi ^\dag }({\gamma ^\mu }{D_\mu } + m)\psi } \right]
\end{equation}
where ${D_\mu } = {\partial _\mu } + i{T^a}A_\mu ^a$ and $m$ is the bare mass. The fermions are in fundamental representation. Through this entire paper we use light-cone gauge (see Appendix \ref{LCgauge} for details). In this gauge convention, $A_-=(A_1-iA_2)/\sqrt{2}=0$, so that the self-interaction term of gauge field vanishes, which greatly simplifies the calculation. The gauge field propagator is given by
\begin{equation}
{G_{ + 3}}(p) =- {G_{3 + }}(p) = \frac{{4\pi i}}{{\kappa {p^ + }}} = 4\pi i\frac{\lambda }{N}\frac{1}{{{p^ + }}},
\end{equation}
where $\lambda= \frac{N}{\kappa }$. At large $N$, $N$ and $\kappa$ are taken to infinity while $\lambda  = \frac{N}{\kappa }$ is held fixed.
Following Ref.~\cite{Bardeen}'s convention of definition,\footnote{Note that the convention of definition for the gap $\Sigma$ and the fermion propagator $S(p)$ that Bardeen~\cite{Bardeen} used is slightly different (though equivalent) from that of Ref.~\cite{Giombi, Frishman2}: Bardeen used definition $\Sigma=\gamma^{\mu}\Sigma_{\mu}$ and thus $S(p)=\frac{1}{{i\slashed{p}+ \Sigma}}$, while the others~\cite{Giombi, Frishman2} used equivalent definition $\Sigma=\gamma^{\mu}\Sigma_{\mu}-m$ and thus the $S(p)=\frac{1}{{i\slashed{p}+ m+\Sigma }}.$ We adopt Bardeen's convention for the consideration of brevity.} the fermion propagator is:
\begin{equation}
\begin{aligned}
S(p) &=\frac{1}{{i\slashed{p}+ \Sigma }}=\frac{1}{{i\slashed{p}+\gamma^{\mu}\Sigma_{\mu}}} 
\end{aligned}
\label{propagator}
\end{equation} 
where we've employed some notation conventions in light-cone gauge:
 \begin{equation}
 \begin{aligned}
 \slashed{p}&={\gamma ^\mu }{p_\mu } ={\gamma ^3 }{p_3 } +{\gamma ^+ }{p_+ } +{\gamma ^- }{p_- }={\gamma _3 }{p^3 } +{\gamma _- }{p^- }+{\gamma_+ }{p^+ } \\ 
p^2&= p_1^2 + p_2^2+p_3^2=p_s^2 +p_3^2=2p^+p^-+p_3^2\\
{\Sigma}& ={\gamma^3}\Sigma_3+{\gamma^+}{\Sigma_+}+{\gamma^-}{\Sigma_-}+{\Sigma_I}
\end{aligned} 
\label{LCconv}
 \end{equation}

At large $N$, planar diagrams in leading order are summed, and we can omit fermion loops when they are in the fundamental representation. Thus the fermion self-energy is determined by the gap equation that includes the summation of ``rainbow'' diagrams
 \begin{eqnarray}
 \Sigma (p) &=& m + \frac{N}{2}\int {\frac{{{d^3}q}}{{{{(2\pi )}^3}}}} \left( {{\gamma ^\mu }S(q){\gamma ^v}} \right){G_{\mu v}}(p - q)  \nonumber \\
&=& {m} - 2i\pi \lambda \int {\frac{{{d^3}q}}{{{{(2\pi )}^3}}}} \frac{1}{{{{(p - q)}^ + }}}\left\{ {{\gamma^3}S(q){\gamma ^+ } - {\gamma ^+ }S(q){\gamma ^3}} \right\}, 
\label{sigmafull}
 \end{eqnarray}
which is exact to leading order at large $N$. 
From the form above, we can see the gap solution is independent of parameter $q_3$. And $\Sigma_3=\Sigma_-=0$ for consistency.\footnote{Note that we also ignored $\delta\Sigma_3$ and $\delta\Sigma_-$ in later curvature study, because our main interest is to give \textbf{sufficient} proof of instability.}
Then the fermion propagator, Eq.~(\ref{propagator}), can be converted into the form:
\begin{equation}
S(p)=\frac{{ - i\slashed{p} - {\gamma ^ + }{\Sigma _ + } + {\Sigma _I}}}{p^2+ M^2},
\end{equation}
where 
 \begin{equation}
 {M^2} = {({\Sigma _I})^2} - 2i{p^ + }{\Sigma _ + }
  \label{physicalMass}
\end{equation} is the induced mass.
Plugging Eq.~(\ref{propagator})$\sim$(\ref{physicalMass}) into Eq.\ (\ref{sigmafull}), it follows\footnote{We can show $M^2$ is momentum-independent (see Appendix.~\ref{M_ind}).}
\begin{equation}
\begin{aligned}
{\Sigma _I}(p) &= {m} - i4\pi \lambda \int {\frac{{{d^3}q}}{{{{(2\pi )}^3}}}} \frac{1}{{{{(p - q)}^ + }}}\frac{{i{q^ + }}}{{q_3^2 + q_s^2 + {M^2}}}  \\
&= {m} + {\lambda }{\sqrt {p_s^2 + {M^2}} } 
\label{sigmaI}
\end{aligned}
\end{equation}
\begin{equation}
\begin{aligned}
{\Sigma _ + }(p) &=&  - i4\pi \lambda \int {\frac{{{d^3}q}}{{{{(2\pi )}^3}}}} \frac{1}{{{{(p - q)}^+}}}\frac{{{\Sigma _I}(q)}}{{q_3^2 + q_s^2 + {M^2}}}\\
&=& - \frac{{i\lambda }}{{2{p^ + }}}\left[ {\lambda p_s^2 + 2m\left( {\sqrt {p_s^2 + {M^2}}- M} \right)} \right].
\end{aligned}
\label{sigmaplus}
\end{equation}
Taking Eq.~(\ref{sigmaI}) and Eq.~(\ref{sigmaplus}) back into Eq.~(\ref{physicalMass}) gives:
\begin{equation}
{M^2} = {\left( {{m} + \lambda M} \right)^2}.
\label{bareTune^2}
\end{equation}
This leads to the relation
\begin{equation}
{m=(1-\lambda)M}.
\label{bareTune}
\end{equation}
Whether the relation is imposed or not may decide the existence of dilaton, the pseudo-Goldstone boson of spontaneous breaking of the scale symmetry, in this theory. In Bardeen's related work~\cite{Bardeen}, a massless dilaton pole in scalar current correlator is identified when this relation Eq.~(\ref{bareTune}) is avoided. In contrast, this relation is adopted by~\cite{Frishman2, Moshe}, and they showed there is no such state. Bardeen himself also noticed this relation will remove the dilaton pole~\cite{letter}~\cite{Bardeen2}, yet he speculated that the source of this relation is merely an artifact from light-cone gauge.\footnote{Note that Ref.~\cite{Moshe} adopted Weyl Gauge, and still showed no such dilaton pole in this theory, if without extra deformation term.} This controversy motivates us to study the (in)stability of the effective potential in this theory.

In the following main body of our work, section~\ref{stab_main}, we shall study the (in)stability around its gap solution (Eq.~(\ref{sigmafull})) by deriving the curvature of the effective potential. In section~\ref{Compare_QCD}, We will compare our (in)stability study with that of QCD-like theories.

\section{(In)stability}
\label{stab_main}

\subsection{General Criteria}
\label{criteria}
The overall scheme of the (in)stability study is taking the second functional variation of the effective potential around the gap solution $\Sigma=\gamma^+\Sigma_++\Sigma_I$ (Eq.~(\ref{sigmafull})) to obtain its curvature. 

In order to make everything be in the real domain, we do the variation over $\overline{\Sigma_+}(q)=i{q^ + }{\Sigma _ +(q) }/{q_s}$ instead of simply over $\Sigma_+(q)$.

The overall variation of effective potential $U$ around the vacuum can be expressed as:
\begin{equation}
{{\delta U}} =\frac{1}{2} \int {\frac{d^2 q}{(2\pi)^2}\frac{d^2 q'}{(2\pi)^2}[} \delta \overline{\Sigma_+}(q),\delta {\Sigma _I}(q)]\left[ {\begin{array}{*{20}{c}}
{\frac{{{\delta ^2}U}}{{\delta\overline{\Sigma_+}(q)\delta \overline{\Sigma_+}(q')}}}&{\frac{{{\delta ^2}U}}{{\delta \overline{\Sigma_+}(q)\delta {\Sigma _I}(q')}}}\\
{\frac{{{\delta ^2}U}}{{\delta \overline{\Sigma_+}(q')\delta {\Sigma _I}(q)}}}&{\frac{{{\delta ^2}U}}{{\delta {\Sigma _I}(q)\delta {\Sigma _I}(q')}}}
\end{array}} \right]\left[ {\begin{array}{*{20}{c}}
{\delta \overline{\Sigma_+}(q')}\\
{\delta {\Sigma _I}(q')}
\end{array}} \right]
\label{overall_vary}
\end{equation}
Note that the integrals on $q_3$ and $q'_3$ have been performed by definition. The matrix
\begin{equation}
\left[ {\begin{array}{*{20}{c}}
{\frac{{{\delta ^2}U}}{{\delta\overline{\Sigma_+}(q)\delta \overline{\Sigma_+}(q')}}}&{\frac{{{\delta ^2}U}}{{\delta \overline{\Sigma_+}(q)\delta {\Sigma _I}(q')}}}\\
{\frac{{{\delta ^2}U}}{{\delta \overline{\Sigma_+}(q')\delta {\Sigma _I}(q)}}}&{\frac{{{\delta ^2}U}}{{\delta {\Sigma _I}(q)\delta {\Sigma _I}(q')}}}
\end{array}} \right]
\label{stabM}
\end{equation}
is the Hessian Matrix spanning in $(\overline{\Sigma_+}, \Sigma_I)$ space. We will call it ``\textbf{stability matrix}".

Next we explain explicitly the methods that we'll use later to argue the saddle-point instability of the effective potential. First we can find ways, as we'll show later concretely, to make the integral diagonalized in momentum space, so that
\begin{equation}
{{\delta U}} =\frac{1}{2}\int {\frac{d^2 q}{(2\pi)^2}\big{\{}a(q)\delta\overline{\Sigma}^2_+(q)+c(q)\delta\Sigma^2_I(q)+ 2 b(q)\delta\overline{\Sigma_+}(q)\delta\Sigma_I(q) }\big{\}}.
\label{acb_U}
\end{equation}

\subsubsection{ Method A}
\label{saddle_1}
If either $a(q)$ or $c(q)$ is non-zero, say $c(q)$, then 
\begin{equation}
{{\delta U}} =\frac{1}{2} \int {\frac{d^2 q}{(2\pi)^2}\frac{1}{c(q)}\big{\{[c(q)\delta{\Sigma_I}+b(q)\delta\overline{\Sigma_+}]^2+\delta\overline{\Sigma_+}^2[a(q)c(q)-b(q)^2]\big{\}}}}
\end{equation}
Thus for the case that $a(q)c(q)-b(q)^2>0$ and $c(q)$ has a definite sign for any $q$, we have $\text{sign}(\delta U)=\text{sign}(c(q))$. Therefore, in this case $c(q)>0$ indicates a local minimum, $c(q)<0$ indicates a local maximum.
For the case that $a(q)c(q)-b(q)^2<0$ and $c(q)$ has a definite sign for any $q$, we prove it is a saddle point by choosing two directions that give opposite sign($\delta U$):
\begin{itemize}
\item Choose direction $\delta\overline{\Sigma_+}=-\frac{c(q)}{b(q)}\delta{\Sigma_I}$, such that ${{\delta U}} =\frac{1}{2} \int {\frac{d^2 q}{(2\pi)^2}\frac{1}{c(q)}\big{\{\delta\overline{\Sigma_+}^2[a(q)c(q)-b(q)^2]\big{\}}}}$. Therefore in this direction $\text{sign}(\delta{U})= -\text{sign}(c(q))$.
\item Choose direction $\delta\overline{\Sigma_+}=0$, such that ${{\delta U}} =\frac{1}{2} \int {\frac{d^2 q}{(2\pi)^2}{c(q)\delta{\Sigma^2_I}}}$. Therefore in this direction $\text{sign}(\delta{U})= \text{sign}(c(q))$.
\end{itemize}
Thus for the case of $a(q)c(q)-b(q)^2<0$ and $c(q)$ (or $a(q)$) has definite sign for any $q$, a saddle point is indicated. We will adopt this as the \textbf{instability criteria} for identifying saddle point in the AF potential case.
Note that this instability criteria is automatically satisfied when either $a(q)$ or $c(q)$ is zero while the other is non-zero for any $q$. When both $a(q)$ and $c(q)$ are zero, it also indicates the saddle-point instability since we can choose $\delta\overline{\Sigma_+}=-\delta{\Sigma_I}$ or $\delta\overline{\Sigma_+}=\delta{\Sigma_I}$ to make the resulting $\delta{U}=\frac{1}{2}\int {\frac{d^2 q}{(2\pi)^2}\big{\{}2 b(q)\delta\overline{\Sigma_+}(q)\delta\Sigma_I(q) }\big{\}}$ give opposite signs.

\subsubsection{ Method B}
\label{saddle_2}
Another way to identify the saddle-point instability is that if we can find different variations of $\delta\overline{\Sigma}_+$ (or $\delta{\Sigma}_I$) that can make $a(q)$ (or $c(q)$) give different signs, we can directly set the other component $\delta{\Sigma}_I$ (or $\delta\overline{\Sigma}_+$) to zero so that the overall potential variation Eq.~(\ref{acb_U}) reduces to
\begin{equation}
{{\delta U}} =\frac{1}{2} \int {\frac{d^2 q}{(2\pi)^2}{a(q)\delta\overline{\Sigma}^2_+}}  \text{                } (\text{or }  {{\delta U}} =\frac{1}{2} \int {\frac{d^2 q}{(2\pi)^2}{c(q)\delta{\Sigma}^2_I}}),
\end{equation}
and thus the sign change of $a(q)$ (or $c(q)$) leads to the sign change of ${\delta U}$ for the chosen variations of $(\delta\overline{\Sigma}_+, 0)$ (or $(0, \delta{\Sigma}_I)$). Then we can directly tell that the potential has the saddle-point instability, even without any knowledge of $b(q)$. This way brings great simplifications when $b(q)$ is very complex to study. We will adopt this way in the case of CJT potential.

\subsection{Auxiliary-Field (AF) effective potential}
\label{VofAction}
In auxiliary field formalism, the effective potential\footnote{For brevity, we normalized it by absorbing the gauge group factor $N$.} in this theory can be derived into the form~\cite{Giombi}:
\begin{equation}
\begin{aligned}
U_{AF}&=U_1+U_2 \\
U_1&= {-   \int {\frac{{{d^3}q}}{{{{(2\pi )}^3}}}} Tr\ln (i\slashed{q} + \Sigma (q)) } \label{pi_2} \\
U_2&= {- \frac{{  1}}{{8\pi i\lambda }}\int {\frac{{{d^2}q}}{{{{(2\pi )}^2}}}} \frac{{{d^2}q'}}{{{{(2\pi )}^2}}}{G^{ - 1}}(q - q')Tr\left({\gamma ^ - }\Sigma (q){\gamma ^3}\Sigma (q')\right)},
\end{aligned}
\end{equation}
where ${G^{ - 1}}(q - q')$ is defined by:
\begin{equation}
\int {\frac{{{d^2}q'}}{{{{(2\pi )}^2}}}} {G^{ - 1}}(q - q')\frac{1}{{{{(q' - p)}^ + }}} = {\left( {2\pi } \right)^2}{\delta ^2}(q - p).
\end{equation}
In the following we derive the explicit form of $G^{-1}$ that will have great importance on our later discussion of stability.
Take derivative on both sides of the equation above:
\begin{eqnarray}
\int {\frac{{{d^2}q'}}{{{{(2\pi )}^2}}}} {G^{ - 1}}(q - q')\frac{\partial}{\partial p_+}\frac{1}{{{{(q' - p)}^ + }}} &=& {\left( {2\pi } \right)^2}\frac{\partial}{\partial p_+}{\delta ^2}(q - p). \nonumber
\end{eqnarray}
Then with the identity
\begin{equation}
 \frac{\partial }{{\partial {p_ + }}}\frac{1}{{{{(p - q)}^ + }}} = 2\pi {\delta ^2}(p - q),
 \end{equation}
we obtain:
 \begin{equation}
  {G^{ - 1}}(q - p)=-2\pi\frac{\partial}{\partial p_+}{\delta ^2}(q - p).
  \label{Ginverse}
 \end{equation}

The first variation of $U_{AF}$ can be easily obtained:
\begin{equation}
\begin{aligned}
{{\delta U_1}} &=-\int \frac{dq_3}{2\pi}Tr (\delta \Sigma(q) \frac{1}{i\slashed{q}+\Sigma(q)})\\
{{\delta U_2}} &=\frac{{-1}}{{8\pi i\lambda }} \int \frac{dq'^2}{(2\pi)^2} G^{-1}(q-q') Tr\left(\delta \Sigma(q) H_{-}(\Sigma(q'))\right),
\end{aligned}
\end{equation}
where we employed short notation convention $H_-(A)=\gamma^3 A \gamma^{-} -\gamma^- A \gamma^3=2(-A_I\gamma^-+A_+I)$ from~\cite{Giombi}.
Next we proceed to their second variation:
\begin{eqnarray}
{{\delta^2 U_1}} &=&\int \frac{dq_3}{2\pi}Tr \left(\delta \Sigma(q) \frac{1}{i\slashed{q}+\Sigma(q)}\delta \Sigma(q')\frac{1}{i\slashed{q}+\Sigma(q)}\right){\delta ^2}(q - q') \nonumber \\
&=&\int \frac{dq_3}{2\pi}Tr(\Delta(q)\delta\Sigma(q')){\delta ^2}(q - q'),
\label{AF_U1_2nd}
\end{eqnarray}
where
\begin{equation}
\Delta(q)=\frac{1}{i\slashed{q}+\Sigma(q)}{\delta\Sigma(q)}\frac{1}{i\slashed{q}+\Sigma(q)}.
\end{equation}
Besides, $\Delta(q)$ can be decomposed into:
\begin{equation}
\Delta(q)=\Delta_I(q)+\Delta_{+}(q)\gamma^{+}+\Delta_{-}(q)\gamma^{-}+\Delta_{3}(q)\gamma^{3}.
\end{equation}
With Eq.~(\ref{S}) and collecting terms in $\gamma$ basis, we obtain
\begin{equation}
\begin{aligned}
\Delta_I(q)&=\big{\{}\delta\Sigma_I{(-q^2+\Sigma_I^2+2i q^+ \Sigma_+)}+\delta\Sigma_+{(-2i q^+ \Sigma_I)}\big{\}}{(q^2+M^2)^{-2}}\\
\Delta_-(q)&=\big{\{}\delta\Sigma_I{(-2i q^+ \Sigma_I)}-\delta\Sigma_+{2 q^{+}q^{+}}\big{\}}{(q^2+M^2)^{-2}}.
\end{aligned}
\label{Delta}
\end{equation}
We don't need the knowledge of $\Delta_+(q)$ and $\Delta_3(q)$ in our later discussions so we don't bother deriving them here. Together with the trace properties of $\gamma$ matrices listed in Appendix (\ref{tracelist}), we have:
\begin{eqnarray}
Tr(\Delta(q)\delta\Sigma(q'))&=&Tr \left((\Delta_I I+\Delta_{\mu} \gamma^{\mu})(\delta \Sigma_I I+\delta\Sigma_{\mu} \gamma^{\mu})\right) \nonumber\\
&=&2(\Delta_I(q) \delta\Sigma_I(q')+\Delta_{-}(q) \delta\Sigma_{+}(q'))\nonumber\\
&=&2\big{\{}\delta\Sigma_I(q)\delta\Sigma_I(q') \left(-q^2+\Sigma_I^2+2i q^+ \Sigma_+\right)-\delta\Sigma_+(q)\delta\Sigma_+(q'){(2 q^{+}q^{+})}\nonumber\\
&+& \left[ \delta\Sigma_+(q)\delta\Sigma_I(q')+\delta\Sigma_+(q')\delta\Sigma_I(q) \right] {(-i 2q^+ \Sigma_I(q))}\big{\}}{(q^2+M^2)^{-2}},
\end{eqnarray}
where in the last step we have substituted Eq.~(\ref{Delta}). Next we substitute above results into Eq.~(\ref{AF_U1_2nd}) and perform the remaining integral on $q_3$, we obtain the stability matrix (see its definition Eq.~(\ref{stabM})) of the potential $U_1$:
\begin{equation}
\left[ \scalemath{1.2}{{\begin{array}{*{30}{c}}
\frac{q_s}{\Sigma_I}&{-1}\\
{-1}&{\frac{(2i{q^ + }{\Sigma _ + }-q_s^2 )}{q_s\Sigma_I}}
\end{array}} }\right]A(q){\delta ^2}(q - q'),
\label{AFstabU1}
\end{equation}
where
\begin{equation}
A(q)=\frac{q_s\Sigma_I}{{(q_s^2+M^2)}^{\frac{3}{2}}}.
\label{Aq}
\end{equation}
Note that for any positive $\lambda$, $\Sigma_I(q)$ is always positive, and then so is $A(q)$.\footnote{Proof: $\Sigma_I(q)=m + {\lambda }{\sqrt{q_s^2 + {M^2}}}=(1-\lambda)M+\lambda\sqrt{q_s^2 + {M^2}}=M+\frac{\lambda q^2_s}{\sqrt{q_s^2+M^2}+M}>0,$ provided $\lambda>0$. And thus
$A(q)=\frac{q_s\Sigma_I}{{(q_s^2+M^2)}^{\frac{3}{2}}}>0.$}

For the second variation of $U_2$:
\begin{equation}
{{\delta^2 U_2}} =\frac{{-1}}{{8\pi i\lambda }} G^{-1}(q-q') Tr(\delta \Sigma(q) H_{-}(\delta\Sigma(q'))),
\end{equation}
where
\begin{eqnarray}
Tr( \delta \Sigma(q) H_-( \delta \Sigma(q')))&=&Tr \left(( \delta \Sigma_I(q) I+ \delta \Sigma_{\mu}(q) \gamma^{\mu})2(- \delta \Sigma_I(q') \gamma^{-} +  \delta \Sigma_{+}(q')I)\right)\nonumber\\
&=&4( \delta \Sigma_I(q)  \delta \Sigma_+(q')- \delta \Sigma_+(q)  \delta \Sigma_I(q')).
\end{eqnarray}
It follows that the stability matrix of the potential $U_2$ is:
\begin{equation}
\left[ {\begin{array}{*{20}{c}}
0&{-\frac{{{q_s}}}{{2\pi \lambda {q^ + }}}{G^{ - 1}}(q - q') }\\
q \leftrightarrow q' &0
\end{array}} \right],
\label{AFstabU2}
\end{equation}
where $q \leftrightarrow q'$ means the lower off-diagonal term is the same as the upper off-diagonal term after exchanging variable $q$ with $q'$.

Combining Eq.~(\ref{AFstabU1}) and Eq.~(\ref{AFstabU2}), we obtain the overall stability matrix of $U_{AF}$:
\begin{equation}
\left[ {\begin{array}{*{20}{c}}
\frac{q_s}{\Sigma_I}A(q){\delta ^2}(q - q')&{-\frac{{{q_s}}}{{2\pi \lambda {q^ + }}}{G^{ - 1}}(q - q') - A(q){\delta ^2}(q - q')}\\
{-\frac{{{q'_s}}}{{2\pi \lambda {q'^ + }}}{G^{ - 1}}(q' - q) - A(q'){\delta ^2}(q - q')}&{\frac{(2i{q^ + }{\Sigma _ + }-q_s^2 )}{q_s\Sigma_I}}A(q){\delta ^2}(q - q'),
\end{array}} \right],
\label{stab_AF}
\end{equation}
Then to study the stability, we follow the method~(\ref{saddle_1}). 
It is obvious that $a(q)=\frac{q_s}{\Sigma_I}A(q)>0$, therefore $U_{AF}$ is stable along $\overline{\Sigma_+}$ direction. The analysis of $c(q)={\frac{(2i{q^ + }{\Sigma _ + }-q_s^2 )}{q_s\Sigma_I}}A(q)$ is given in Appendix.~\ref{cq} and is shown to have definite sign for any momentum for any positive $\lambda$.
In order to obtain the $b(q)$, we need to diagonalize the non-local $G^{-1}(q-q')$ term in momentum space:
note that from Eq.~(\ref{Ginverse}), after relabeling:
\begin{equation}
  {G^{ - 1}}(q - q')=-2\pi\frac{\partial}{\partial q'_+}{\delta ^2}(q - q')=2\pi\frac{\partial}{\partial q_+}{\delta ^2}(q - q')=2\pi\frac{\partial q_s}{\partial q_+}\frac{\partial}{\partial q_s}{\delta ^2}(q - q')=2\pi \frac{q^+}{q_s}\frac{\partial}{\partial q_s}{\delta ^2}(q - q').
 \end{equation}Substitute this into the the non-local term in the stability matrix (\ref{stab_AF}) to make it diagonalized in momentum space:
\begin{equation}
\begin{aligned}
-\frac{1}{2\pi \lambda}\int \frac{d^2 q}{(2\pi)^2}\frac{d^2 q'}{(2\pi)^2} \delta{\overline{\Sigma _ + }(q)} \frac{q_s}{q^+}G^{ - 1}(q - q') \delta {\Sigma _I}(q') &=-\frac{1}{\lambda} \int \frac{d^2 q}{(2\pi)^2}\frac{d^2 q'}{(2\pi)^2} {\delta{\overline{\Sigma _ + }(q)}} \frac{\partial}{\partial q_s}{\delta ^2}(q - q') \delta {\Sigma _I}(q')\\&=-\frac{1}{\lambda} \int \frac{d^2 q}{(2\pi)^2}{\delta{\overline{\Sigma _ + }(q)}} \frac{\partial}{\partial q_s} \delta {\Sigma _I}(q).
\end{aligned}
\label{GinvT}
\end{equation}
Inspecting~(\ref{GinvT}), we can pick the variation function $\delta {\Sigma_I}$ which satisfies
\begin{equation}
\frac{1}{\lambda} \frac{\partial}{\partial q_s} \delta {\Sigma _I}(q)=-\epsilon A(q) \delta {\Sigma _I}(q)
\label{variation_pick}
\end{equation}
where $\epsilon$ is arbitrary variable, yet the convergence of $\delta{\Sigma_I}$ require it to be non-negative.\footnote{Eq.~(\ref{variation_pick}) can be easily solved to obtain: $$\delta{\Sigma _I}(q) = \delta{\Sigma _I}(0) \text{ }\left({(\frac{q_s}{M})^2+1}\right)^{-\frac{\lambda^2\epsilon}{2}}exp\left(\lambda \epsilon (\frac{m}{\sqrt{q_s^2+M^2}}-\frac{m}{M})\right).$$ We can see it is always convergent for $\epsilon\geq0$.} Then the stability matrix~(\ref{stab_AF}) reduces to
\begin{equation}
\left[ \scalemath{1.2}{{\begin{array}{*{30}{c}}
\frac{q_s}{\Sigma_I}&{\epsilon-1}\\
{\epsilon-1}&{\frac{(2i{q^ + }{\Sigma _ +}-q_s^2 )}{q_s\Sigma_I}}
\end{array}} }\right]A(q){\delta ^2}(q - q'),
\label{AFstabM2}
\end{equation}
so that $b(q)=({\epsilon-1})A(q)$.
Therefore, following method~(\ref{saddle_1}), the instability criteria:
\begin{equation}
\begin{aligned}
a(q)c(q)-b(q)^2&=\left(\frac{q_s}{\Sigma_I}{\frac{(2i{q^ + }{\Sigma _ + }-q_s^2 )}{q_s\Sigma_I}}-(\epsilon-1)^2\right)A^2(q)\\
&=-\frac{q_s^2}{(q_s^2+M^2)^2}-(\epsilon^2-2\epsilon)A^2(q)<0,\\
\end{aligned}
\label{AF_acb}
\end{equation} 
for the variation $\delta{\Sigma_I}$ that satisfying Eq.~(\ref{variation_pick}) with $\epsilon>2$ or $\epsilon=0$. Besides, $c(q)$ has definite sign for any momentum for any positive $\lambda$ (see Appendix.~\ref{cq}).
Therefore, following method~(\ref{saddle_1}), we've proved the gap solution is a saddle point of the AF effective potential. This conclusion is independent of the value of coupling $\lambda$ (as long as $\lambda$ is positive).
\subsection{Cornwall-Jackiw-Tomboulis (CJT) effective potential}
\label{sec_CJT}
Since the general CJT effective potential\footnote{For brevity, we normalized it by absorbing the gauge group factor $N$.} has following structure~\cite{CJT, Haymaker}:
\begin{equation}
U_{CJT}=i\text{}\Tr (\ln S^{-1}+S_0^{-1}S-\frac{1}{2}SGS),
\end{equation}
where $S_0$ and $S$ is the free and full fermion propagator respectively and $G$ is the gauge propagator. This inspires us to propose that the explicit form of the CJT effective potential in this theory is
\begin{equation}
\begin{aligned}
{U_{CJT}}&=U_1+U_2 \\
U_1&= {- \int {\frac{{{d^3}q}}{{{{(2\pi )}^3}}}} Tr [\ln (i\slashed{q} + \Sigma (q)) }+\frac{i\slashed{q}+m }{i\slashed{q} + \Sigma (q)} ] \\ 
U_2&= 2i\pi \lambda{\int {\frac{{{d^3}q}}{{{{(2\pi )}^3}}}} \frac{{{d^3}q'}}{{{{(2\pi )}^3}}} \frac{1}{(q - q')^+} Tr\left(S(q){\gamma ^ + }S(q'){\gamma ^3} \right)},
\end{aligned}
\end{equation}
where $S=({{i\slashed{q} + \Sigma}})^{-1}$ and $S_0=({{i\slashed{q}+m }})^{-1},$ and $m$ is the bare mass. We derive the gap equation from above potential, which also provides a double check on its correctness:
\begin{eqnarray}
{\delta U_1}&=&-\int \frac{dq_3}{2\pi}Tr \left(\delta \Sigma(q) (\frac{1}{i\slashed{q}+\Sigma(q)}-\frac{1}{i\slashed{q}+\Sigma(q)}+\frac{1}{i\slashed{q}+\Sigma(q)}{(\Sigma(q)-m)}\frac{1}{i\slashed{q}+\Sigma(q))}\right)\nonumber\\
&=&-\int \frac{dq_3}{2\pi}Tr\left(\Delta(q)(\Sigma(q)-m)\right),\\
{\delta U_2}&=&2i\pi \lambda \int \frac{dq_3}{2\pi} \int \frac{{d^3}q'}{(2\pi)^3} \frac{1}{(q - q')^+}Tr\left(\delta S(q)(\gamma^3 S(q') \gamma^{+} -\gamma^+ S(q') \gamma^3)\right)\nonumber\\
&=&-2i\pi \lambda \int \frac{dq_3}{2\pi} \int \frac{{d^3}q'}{(2\pi)^3}  \frac{1}{(q - q')^+} Tr(\Delta(q) H_+(S(q')),
\label{1stU2}
\end{eqnarray}
where in second line we used the fact $\delta S=-\frac{1}{i\slashed{q}+\Sigma(q)}{\delta\Sigma(q)}\frac{1}{i\slashed{q}+\Sigma(q)}=-\Delta(q)$ and short notation convention $H_+(A)=\gamma^3 A \gamma^{+} -\gamma^+ A \gamma^3=2(A_I\gamma^+-A_-I)$.
Therefore
\begin{equation}
{\delta U_{CJT}}=-\int \frac{dq_3}{2\pi}Tr\left(\Delta(q)\left(\Sigma(q)-m+2i\pi \lambda \int \frac{{d^3}q'}{(2\pi)^3}  \frac{1}{(q - q')^+} H_+(S(q')\right) \right).
\label{1st_var_CJT}
\end{equation}
It is obvious that ${\delta U}$=0 gives exactly the right gap equation (Eq.~(\ref{sigmafull})). To study the (in)stability around the gap $\Sigma$, we take the second functional variation:\footnote{Note that we can drop the variation over $\Delta(q)$ since its multiplicative factor vanishes at the gap, observing from Eq.~(\ref{1st_var_CJT}).}
\begin{equation}
{\delta^2 U_1}=-\int \frac{dq_3}{2\pi}Tr(\Delta(q)\delta\Sigma(q'))\delta^2(q-q')
\label{2ndU1}
\end{equation}
Note that this has the exactly same form as the Auxiliary field version Eq.~(\ref{AF_U1_2nd}), except with the opposite overall sign. Thus for the stability matrix of $U_1$, we can directly borrow the result Eq.~(\ref{AFstabU1}), except with a flipped sign:
\begin{equation}
-\left[ \scalemath{1.2}{{\begin{array}{*{30}{c}}
\frac{q_s}{\Sigma_I}&{-1}\\
{-1}&{\frac{(2i{q^ + }{\Sigma _ + }-q_s^2 )}{q_s\Sigma_I}}
\end{array}} }\right]A(q){\delta ^2}(q - q'),
\end{equation}

As to the second variation of $U_2$, from Eq.~(\ref{1stU2}) we have:
\begin{eqnarray}
{\delta^2 U_2}&=&-2i\pi \lambda \int \frac{dq_3}{2\pi}\int \frac{{d}q'_3}{2\pi} \frac{1}{(q - q')^+}Tr(\Delta(q) H_+(\delta S(q')))  \nonumber\\
&=&2i\pi \lambda\int \frac{dq_3}{2\pi}\int \frac{{d}q'_3}{2\pi} \frac{1}{(q - q')^+}Tr(\Delta(q) H_+(\Delta(q'))),
\label{2ndU2}
\end{eqnarray}
with
\begin{eqnarray}
Tr\left(\Delta(q) H_+(\Delta(q'))\right)&=&Tr \left((\Delta_I I+\Delta_{\mu} \gamma^{\mu})2(\Delta_I \gamma^{+} - \Delta_{-}I)\right)\nonumber\\
&=&4\left(-\Delta_I(q) \Delta_-(q')+\Delta_-(q) \Delta_I(q')\right),
\end{eqnarray}
where
\begin{eqnarray}
 \Delta_I(q) \Delta_-(q')&=&\big{\{}\delta \Sigma_I(q) \delta\Sigma_I(q')(-q^2+\Sigma_I^2+2i q^+ \Sigma_+)(-2i q'_-\Sigma_I(q'))\nonumber\\
&+&\delta\Sigma_+(q)\delta\Sigma_I(q'){(-4q^+q'^+\Sigma_I(q)\Sigma_I(q'))}+\delta\Sigma_I(q)\delta\Sigma_{+}(q')(-2q'^{+}q'^{+}(-q^2+\Sigma_I^2+2i q^+ \Sigma_+))\nonumber\\
&+&\delta\Sigma_+(q)\delta\Sigma_+(q')(4i q^+\Sigma_I(q)q'^{+}q'^{+})\big{\}}\left((q^2+M^2)(q'^2+M^2)\right)^{-2}
\end{eqnarray}
Substitute this into Eq.~(\ref{2ndU2}) and perform the remaining integral on $q_3$, we obtain the stability matrix of the potential $U_2$:
\begin{equation}
2\pi\lambda \frac{1}{(q-q')^+}\frac{q'^+}{q'_s}A(q)A(q')\left[ \scalemath{1}{{\begin{array}{*{30}{c}}
-2\frac{q'_s}{\Sigma_I(q')}&1-\frac{q'_s q^+}{q'^+ q_s}\frac{q_s}{\Sigma_I(q)}\frac{(2i{q'^ + }{\Sigma _ + }-q'^2_s )}{q'_s\Sigma_I(q')}\\
q \leftrightarrow q'&-2\frac{(2i{q^ + }{\Sigma _ + }-q_s^2 )}{q_s\Sigma_I(q)}
\end{array}}}\right]
\label{CJTstabM2}
\end{equation}
where $A(q)$ was defined in Eq.~(\ref{Aq}). Therefore the overall stability matrix of $U_{CJT}$ is:
\begin{equation}
\left[ \scalemath{1}{{\begin{array}{*{30}{c}}
\frac{q'_s}{\Sigma_I(q')}K(q,q')&b(q,q')\\
b(q',q)&\frac{(2i{q^ + }{\Sigma _ + }-q_s^2 )}{q_s\Sigma_I}K(q,q')
\end{array}} }\right]A(q),
\label{stab_CJT}
\end{equation}
where $K(q,q')=-\delta^2(q-q')-4\pi\lambda \frac{1}{(q-q')^+}\frac{q'^+}{q'_s}A(q')$, and $b(q,q')=\delta^2(q-q')+ \frac{2\pi\lambda}{(q-q')^+}\frac{q'^+}{q'_s}A(q')(1-\frac{q'_s q^+}{q'^+ q_s }\frac{q_s}{\Sigma_I(q)}\frac{(2i{q'^ + }{\Sigma _ + }-q'^2_s )}{q'_s\Sigma_I(q')}).$ 

We choose the second way of arguing saddle-point instability to avoid the involvement of $b(q,q')$, following method~(\ref{saddle_2}).
For the first diagonal element ${\frac{{{\delta ^2}U}}{{\delta\overline{\Sigma_+}(q)\delta \overline{\Sigma_+}(q')}}}$ of stability matrix~(\ref{stab_CJT}), the full integral form is
\begin{equation}
\int {\frac{d^2 q}{(2\pi)^2}{\frac{d^2 q'}{(2\pi)^2}\delta \overline{\Sigma_+}(q)}{\frac{{{\delta ^2}U}}{{\delta\overline{\Sigma_+}(q)\delta \overline{\Sigma_+}(q')}}}\delta \overline{\Sigma_+}(q')}=
\int {\frac{d^2 q}{(2\pi)^2}\delta (\overline{\Sigma_+}(q)) A(q)\int \frac{d^2 q'}{(2\pi)^2}} K(q,q')\frac{q'_s}{\Sigma_I(q')}\delta (\overline{\Sigma_+}(q'))\
\label{CJT_11}
\end{equation}
To diagonalize it in momentum space, we choose the variation $\delta\overline{\Sigma_+}$ so that
\begin{equation}
\int \frac{d^2 q'}{(2\pi)^2} K(q,q')\frac{q'_s}{\Sigma_I(q')}\delta \overline{\Sigma_+}(q')=-\eta \frac{q_s}{\Sigma_I(q)}\delta \overline{\Sigma_+}(q)
\label{K_int}
\end{equation}
where $\eta$ is arbitrary real parameter, yet the convergence of $\delta \overline{\Sigma_+}(q)$ requires ${\eta}<1$.\footnote{Solution of Eq.~(\ref{K_int}): denote $\phi(q)=\frac{q_s}{\Sigma_I(q)}\delta \overline{\Sigma_+}(q)$.
Using identity (\ref{angular2}) while doing the integral, Eq.~(\ref{K_int}) gives $-\phi(q)+2\lambda\int dq'_s\theta(q'_s-q_s)A(q')\phi(q')=-\eta \phi_(q).$
Differentiate respects to $q_s$, we obtain: $(1-\eta)\phi'(q)=-2\lambda A(q)\phi(q)$, the solution of which is: $\phi(q) \sim \text{ }\left({{q_s}^2+M^2}\right)^{-\frac{\lambda^2}{1-\eta}}exp\left(\frac{2\lambda}{1-\eta}  (\frac{m}{\sqrt{q_s^2+M^2}})\right)$, up to dimensional normalization. We can also see the convergence of $\delta \overline{\Sigma_+}(q)$ requires ${\eta}<1$.} Then 
\begin{equation}
a(q)={\frac{{{\delta ^2}U}}{{\delta\overline{\Sigma_+}(q)\delta \overline{\Sigma_+}(q)}}}=-\eta A(q)\frac{q_s}{\Sigma_I(q)}=-\eta\frac{q^2_s}{{(q_s^2+M^2)}^{\frac{3}{2}}} \leq0
\end{equation}
for $\eta\geq0$.
This signals instability with the chosen variation $\delta\overline{\Sigma_+}(q)$. Without ruining the convergence, one can also change the variation with $\eta \rightarrow -\eta$, so that $a(q)\geq0$. Thus we manage to identify a saddle-point instability, following method~(\ref{saddle_2}). 

Similarly, for the other diagonal element ${\frac{{{\delta ^2}U_{CJT}}}{{\delta{\Sigma_I}(q)\delta {\Sigma_I}(q')}}} \sim \frac{(2i{q^ + }{\Sigma _ + }-q_s^2 )}{q_s\Sigma_I}K(q,q')$, the factor $K(q,q')$ can also just contribute either a negative sign or a positive sign, depending on the variation $\delta{\Sigma_I}(q)$ chosen, which makes $c(q)={\frac{{{\delta ^2}U_{CJT}}}{\delta{\Sigma_I}\delta {\Sigma_I}}}$ has the opposite or same sign comparing with its AF version Eq.~(\ref{c_AF}). Especially, it vanishes in the spontaneous symmetry breaking limit ($\lambda=1$). 

Therefore, in general, the gap solution is a saddle point of the CJT effective potential. 

For the massless fermion phase, it's obvious that all the previous conclusions of instability still hold after taking $M\to0$.

\section{Comparison with QCD-like theories}
\label{Compare_QCD}
In this section we compare the (in)stability of this theory with that of Landau-gauge QCD-like theories. In Landau gauge, the angular integral for the Dirac-vector component of the self-energy vanishes so only a singlet $\Sigma=\Sigma\delta^{\alpha\beta}$ component is considered. In the stability studies of the Landau-gauge QCD-like theories~\cite{Haymaker}, the gluon self-interactions are usually ignored for simplification, since their main interest is the dynamical chiral symmetry breaking. Ref.~\cite{Haymaker} studied the stability of 4D QCD in Landau gauge. They argued that around the gap solution, the AF potential is always stable while the CJT potential has the saddle-point instability, in contrast to our study of 3D Chern-Simons theory with fermions at large $N$, where both AF and CJT has the saddle-point instabilities at the gap solution. To see the reason of this difference, note that though both this theory and QCD-like theories have similar structures for the CJT potential~\cite{Haymaker}:
\begin{equation}
U_{CJT} = i\text{}\Tr (\ln S^{-1}+S_0^{-1}S-\frac{1}{2}SGS),
\label{U_CJT_general}
\end{equation}
and the AF potential
\begin{equation}
U_{AF} = i\text{}\Tr \left(\ln S^{-1}-\frac{1}{2}\Sigma G^{-1}\Sigma \right),
\label{U_AF_general}
\end{equation}
yet the form of the non-local part $G$ (or $G^{-1}$) and math structure of $\gamma$ matrices are different in different theories in different dimensions, and thus lead to different form of gap solutions and different results of (in)stability.
More explicitly, the curvature for the singlet gap component in the general CJT potential Eq.~(\ref{U_CJT_general}), after doing angular integral, has the structure:
\begin{equation}
\begin{aligned}
\frac{\delta^2 U_{CJT}}{\delta\Sigma{(q)}\Sigma{(q')}} &\sim \tilde{A}(q) \delta(q-q')- \tilde{A}(q) \tilde{M}(q,q') \tilde{A}(q')\\
&=\tilde{A}(q)[\delta(q-q')- \tilde{M}(q,q') \tilde{A}(q')],
\end{aligned}
\label{curv_CJT}
\end{equation}
where we've used tilde hat to avoid notation conflict, and ``$\sim$" indicates that we ignore any positive constant factor (like phase factor $\frac{1}{2\pi}$). $\tilde{A}(q)$ is the curvature of the local part of the potential, and $\tilde{M}(q,q')$ includes the non-local part.
In 4D QCD, 
\begin{equation}
\tilde{A}(q)_{QCD}\sim q^3 \frac{q^2-\Sigma^2(q)}{(q^2+\Sigma^2(q))^2},
\end{equation}
which would change sign over momentum at their gap solution. 
And 
\begin{equation}
\tilde{M}(q,q')_{QCD}\sim \frac{g^2(q^2)}{q^2}\theta(q-q')+\frac{g^2(q'^2)}{q'^2}\theta(q'-q),
\end{equation}
which is definitely positive for any momentum, thus gives definite negative contribution to the curvature Eq.~(\ref{curv_CJT}). Thus Ref.~\cite{Haymaker} can show the instability by choosing the step function as variation $\delta \Sigma=\theta(p_0-p)$ where $p_0$ is the point below which $\tilde{A}(p)$ turns negative.

However, these curvature terms' sign behaviours are totally different in our case. Take the $\overline{\Sigma_+}$ component for example, referring to Eq.~(\ref{stab_CJT}) and do conversion to radial coordinates:$\int\frac{d^2q}{(2\pi)^2} \to \int \frac{dq_s q_s}{(2\pi)^2}\int d\theta$, then we have
 \begin{equation}
 \tilde{A}(q)_{CS}\sim-\frac{q_s^2}{\Sigma_I}A(q)=-\frac{q_s^3}{{(q_s^2+M^2)}^{\frac{3}{2}}},
 \end{equation}
 which is definitely negative for any momentum.
Then to match the structure of Eq.~(\ref{curv_CJT}), we have
\begin{equation}
\tilde{M}(q,q')_{CS}\sim -\frac{\lambda\Sigma_I(q')}{q_s q'_s}\theta(q_s-q'_s)-\frac{\lambda\Sigma_I(q)}{q_s q'_s}\theta(q'_s-q_s),
\end{equation}
which is also definitely negative for any momentum. Therefore, the sign behaviours of both $\tilde{A}(q)$ and $\tilde{M}(p,q)$ are totally different than those of the 4D QCD case. Thus, to argue the total curvature Eq.~(\ref{curv_CJT}), we shouldn't follow their method that choosing step function as variations as the 4D QCD case above. Instead, in Eq.~(\ref{K_int}), we chosen convergent variations so that $[\delta(q-q')- \tilde{M}(q,q') \tilde{A}(q')]$ gives definite sign after diagonalization in momentum space (integrate out $q'$). Unlike the 4D QCD~\cite{Haymaker}, the diagonalization (in momentum space) can be done purely analytically and can guarantee convergence here due to the exact solubility of this theory.

For the AF potential Eq.~(\ref{U_AF_general}), the curvature has following structure:
\begin{equation}
\begin{aligned}
\frac{\delta^2 U_{AF}}{\delta\Sigma{(q)}\Sigma{(q')}} &\sim q^{D-1}\tilde{M}^{-1}(q,q')q'^{D-1} -\tilde{A}(q) \delta(q-q').
\end{aligned}
\label{curv_AF}
\end{equation}
where $D$ denotes the dimension number. The component decomposition $\Sigma=\gamma^+\Sigma_++\Sigma_I$ in our case, makes our study very different from the QCD-like theories~\cite{Haymaker}, where only one singlet component $\Sigma=\Sigma\delta^{\alpha\beta}$ was studied. 
For example, in our case the curvature form Eq.~(\ref{curv_AF}) only appears in the off-diagonal element of stability matrix Eq.~(\ref{stab_AF}), due to the properties of $\gamma$ matrices associated with the $\Sigma$ decomposition. Moreover, because of the two-component space $(\overline{\Sigma_+}, \Sigma_I)$, we can use method~(\ref{saddle_1}) to argue the saddle-point instability. Finally, as we've shown under Eq.~(\ref{stab_AF}), with the exact solubility of this theory, we can deal with everything purely analytically, even for the nonlocal $\tilde{M}^{-1}(q,q')$ part, so that we don't need to resort to numerical methods to argue the (in)stability like the 4D QCD case studied in Ref.~\cite{Haymaker}.

All in all, different aspects like the modifications on the curvature functions' sign behaviours over momentum, the extension to two-component space $(\overline{\Sigma_+}, \Sigma_I)$ rather than one singlet $\Sigma_I$, and the exact solubility, make our analyses and conclusions of this theory very different than those of the QCD-like theories~\cite{Haymaker}.
\section{Summary and Discussions}
In this paper we have shown the instability of the 3D $U(N)$ Chern-Simons gauge theory with fundamental fermions at large $N$ around its gap solution $\Sigma=\gamma^+\Sigma _ ++\Sigma_I $. Both the AF and CJT effective potentials are studied and the saddle-point instabilities are shown in both potentials for any positive coupling $\lambda$ at their gap equation. These instabilities hold for both massless ($M=0$) and massive ($M\neq0$) fermion phases. Finally, a comparison with the QCD-like theories is given.

There are still some open questions to explore: 
\begin{itemize}
\item We studied the most widely used effective potentials: AF and CJT. Is there any other effective potential for this theory that can make the gap solution stable? If not, is this theory ``sick" due to this saddle-point instability, just like those higher derivative theories are ``sick'' due to their Ostrogradsky instability? What is the fundamental source of this ``sickness''? Is it coming from the large $N$ taken? Or from the Chern-Simons theory's topological or conformal features?
\item It is also interesting to explore this saddle-point instability's implications on the bosonic and the supersymmetric Chern-Simons theory, the holographic dual and related finite temperature studies~\cite{Giombi}~\cite{ CS_Boson, CS_SUSY, CS_Holo, CS_Thermal}.
\end{itemize}
\begin{acknowledgments}
This work occurred under the supervision of Prof.~Bob Holdom. I am grateful for his kind and incisive guidance and for many helpful discussions. I thank Prof. W.A Bardeen very much for communications in the early stage of this work. This research was supported in part by the Natural Sciences and Engineering Research Council of Canada.
\end{acknowledgments}

\appendix
\appendixpage
\addappheadtotoc
\section{Light-cone Gauge Basics}
\label{LCgauge}
In light-cone gauge there are such definition convention\cite{Giombi}:
\begin{equation}
\begin{aligned}
{x^ \pm } &= \frac{1}{{\sqrt 2 }}({x^1} \pm i{x^2}) \\
{A^ \pm } &= {A_ \mp } = \frac{1}{{\sqrt 2 }}({A^1} \pm i{A^2})  \\
{p^ \pm } &= {p_ \mp } = \frac{1}{{\sqrt 2 }}({p^1} \pm i{p^2})  \\
p_s^2 &= p_1^2 + p_2^2 = 2{p^ + }{p^ - }\\
p^2&= p_1^2 + p_2^2+p_3^2=p_s^2 +p_3^2
\end{aligned}
\end{equation}
In light-cone gauge, the gamma matrices satisfy
\begin{equation}
\{ {\gamma ^\mu },{\gamma ^\nu }\}  = 2{g^{\mu \nu }}
\label{gamma}
\end{equation}
where ${g^{ +  + }} = {g^{ -  - }} = 0$ and ${g^{ +  - }} = {g^{ -  + }} = 1$.
With ${\gamma ^3} = {\gamma ^ + }{\gamma ^ - } - 1$, it follows that ${({\gamma ^3})^2} = I$ and $\{ {\gamma ^ \pm },{\gamma ^3}\}  = 0$.
Here we list following relations which can be easily derived from above definitions.
\begin{equation}
\begin{array}{lr}
Tr({\gamma ^ + }{\gamma ^ - }) = 2  \\
Tr({\gamma ^\pm}) = 0 \\
Tr({\gamma ^3}) = 0 \\
Tr({\gamma ^ \pm }{\gamma ^3}) = 0 \\
Tr({\gamma ^ + }{\gamma ^ - }{\gamma ^3}) = 2 \\
Tr({\gamma ^ - }{\gamma ^ + }{\gamma ^3}) = -2 
\label{tracelist}
\end{array} 
\end{equation} 
From the metric convention, we see that we can lift or lower the `+' (`-') index to its opposite `-' (`+') without changing its value. \\ 
For example, 
\begin{equation}
\begin{aligned}
{\gamma ^ + } &= {\gamma _ - };  {\gamma ^ - } = {\gamma _ + }\\
{\Sigma ^ - } &= {\Sigma _ + } ;  {\Sigma ^ + } = {\Sigma _ - }  \\
{p ^ +}&={p_-}  ;  {p ^ -}={p_+},
\end{aligned}
\end{equation}
and $A^3=A_3$ for any variable $A$.
It follows from above that~\cite{Bardeen}:
\begin{eqnarray}
\frac{1}{{i\slashed{q} + \Sigma }} &=& \frac{1}{{i[{\gamma ^3}{q_3} + {\gamma ^ + }({q_ + } + {\Sigma _ + }/i) + {\gamma ^ - }{q_ - }] + {\Sigma _I}}} \nonumber \\
&=&\frac{{ - i[{\gamma _3}{q^3} + {\gamma ^ - }{q_ - } + {\gamma ^ + }({q_ + } + {\Sigma _ + }/i)] + {\Sigma _I}}}{{{{({q_3})}^2} + ({\gamma ^ + }{\gamma ^ - } + {\gamma ^ - }{\gamma ^ + })({q_ + }{q_ - } + {q^ + }{\Sigma _ + }/i) + {{({\Sigma _I})}^2}}} \nonumber \\
&=&\frac{{ - i\slashed{q} - {\gamma ^ + }{\Sigma _ + } + {\Sigma _I}}}{q^2+ {M^2}}.
\label{S}
\end{eqnarray}
And its square:
\begin{eqnarray}
\frac{1}{(i\slashed{q} + \Sigma )^2} &=&\frac{({ - i\slashed{q} - {\gamma ^ + }{\Sigma _ + } + {\Sigma _I}})^2}{(q^2+ {M^2})^2} \nonumber \\
&=&\frac{(-q^2+\Sigma_I^2+2i q^+ \Sigma_+)-2(\gamma^+\Sigma_++ i\slashed{q})\Sigma_I}{(q^2+ {M^2})^2}
\label{Ssquare}
\end{eqnarray}
As to the angular integral, it's easy to prove~\cite{Giombi}:
\begin{eqnarray}
\int_0^{2\pi } {d\theta } \frac{{{q^ + }}}{{{{(p - q)}^ + }}} &=& - 2\pi \theta (q_s - p_s),
\label{angular2}
\end{eqnarray}
and
\begin{equation}
\int_0^{2\pi } {d\theta } \frac{1}{{{{(p - q)}^ + }}} = \frac{1}{{{p^ + }}}2\pi \theta (p_s - q_s),
\end{equation}
where $\theta (p_s - q_s)$ is the Heaviside step function. Then it follows that:
\begin{equation}
 \frac{\partial }{\partial {p_ + }}\frac{1}{{{{(p - q)}^ + }}} = 2\pi {\delta ^2}(p - q)
 \label{dr_iden}
\end{equation}
\section{Proof of $\frac{d M^{2}}{d p}=0$}
\label{M_ind}
Here we give a simple proof that the induced mass $M$ is momentum-independent by showing its derivative over momentum is zero. To the author's best knowledge, I haven't found any such proof anywhere. So I derived it here for completion. Refer to the definition of $M$, Eq.~(\ref{physicalMass}):
\begin{equation*}
 {M^2} = {({\Sigma _I})^2} - 2i{p^ + }{\Sigma _ + }
\end{equation*}
Taking derivative:
\begin{eqnarray}
\frac{{\partial {M^2}}}{{\partial {p_ + }}} &=& \frac{{\partial \Sigma _I^2}}{{\partial {p_ + }}} - 2i{p^+ }\frac{{\partial {\Sigma _ + }}}{{\partial {p_ + }}}\\
 &=& 2{\Sigma _I}\frac{{\partial {\Sigma _I}}}{{\partial {p_ + }}} - 2i{p^+ }\frac{{\partial {\Sigma _ + }}}{{\partial {p_ + }}} \label{dM}
\end{eqnarray}
Taking derivative from Eq.~(\ref{sigmaI}) and Eq.~(\ref{sigmaplus}), we have: 
\begin{eqnarray*}
\frac{{\partial {\Sigma _I}(p)}}{{\partial {p_ + }}} &=&  - i4\pi \lambda \int {\frac{{{d^3}q}}{{{{(2\pi )}^3}}}} 2\pi {\delta ^2}(p - q)\frac{{i{q^ + }}}{{q_3^2 + q_s^2 + {M^2}}}\\
 &=& - i4\pi \lambda \int {\frac{{d{q_3}}}{{{{(2\pi )}^2}}}} \frac{{i{p^ + }}}{{q_3^2 + p_s^2 + {M^2}}}\\
 \frac{{\partial {\Sigma _ + }(p)}}{{\partial {p_ + }}}{\rm{ }} &=&  - i4\pi \lambda \int {\frac{{{d^3}q}}{{{{(2\pi )}^3}}}2\pi } {\delta ^2}(p - q)\frac{{{\Sigma _I}(q)}}{{q_3^2 + q_s^2 + {M^2}}}\\
  &=&  - i4\pi \lambda \int {\frac{{d{q_3}}}{{{{(2\pi )}^2}}}} \frac{{{\Sigma _I}(p)}}{{q_3^2 + p_s^2 + {M^2}}}
\end{eqnarray*}
Plugging them back to Eq.~(\ref{dM}), without the need of further integration, we can easily observe those two terms have only difference of opposite sign, thus they cancel each other. Therefore we have 
\begin{equation*}
\frac{{\partial {M^2}}}{{\partial {p_ + }}} = 0
\end{equation*}
Take conjugation, one obtain:
\begin{equation*}
\frac{{\partial {M^2}}}{{\partial {p_ - }}} = 0.
\end{equation*}
Obviously
\begin{equation*}
\frac{{\partial {M^2}}}{{\partial {p_ s }}} = 0.
\end{equation*}

\section{Sign of $c(q)$} 
\label{cq}
Here we give the study on the $c(q)=\frac{{{\delta ^2}U}}{{\delta {\Sigma _I}(q)\delta {\Sigma _I}(q)}}$ of the AF potential, which has the interesting $\lambda$-dependent and momentum-independent sign behaviour. 
\begin{equation}
\begin{aligned}
c(q) &= {\frac{(2i{q^ + }{\Sigma _ + }-q_s^2 )}{q_s\Sigma_I}}A(q)\\
&= \frac{A(q)}{q_s\Sigma_I}\left[{{2\lambda m(\sqrt {q_s^2 + {M^2}} - M) - (1 - {\lambda ^2})q_s^2}}\right]\\
&=\lambda(1-\lambda)\frac{A(q)q_s}{\Sigma_I}\left[\frac{2}{\sqrt{1+(\frac{q_s}{M})^2}+1}-(1+\frac{1}{\lambda})\right],
\end{aligned}
\label{c_AF}
\end{equation}
where in second line we substituted the gap solution Eq.~(\ref{sigmaI}), and the relation Eq.~(\ref{bareTune}). Since $\frac{2}{\sqrt{1+({q_s}/{M})^2}+1} \in (0, 1]$, therefore
$\left[\frac{2}{\sqrt{1+({q_s}/{M})^2}+1}-(1+\frac{1}{\lambda})\right]<0$ for any positive $\lambda$. Thus
\begin{itemize}
 \item For $\lambda>1$, we have $c(q)>0$, which means $U_{AF}$ is stable along $\Sigma_I$ direction
 \item For $0\leq\lambda<1$, we have $c(q)<0$, which means $U_{AF}$ is unstable along $\Sigma_I$ direction. Yet we've shown it is stable along $\overline{\Sigma_+}$ direction ($a(q)>0$), thus a saddle point is identified in this case.
\item For $\lambda=1$, we have $c(q)=0$, thus the instability criteria $a(q)c(q)-b(q)^2<0$ and $a(q)$ has definite sign over any momentum, referring to the method~(\ref{saddle_1}), this indicates the saddle-point instability. Besides, from Eq.~(\ref{bareTune}) we know $\lambda=1$ gives $m=0$ but $M$ can be non-zero. Thus this limit corresponds to the spontaneous scale symmetry breaking limit without any explicit breaking. Therefore we can conclude that the dynamical massive phase caused solely by spontaneous scale symmetry breaking is also not stable.
\end{itemize}
Note that this $\lambda$ dependence is removed in saddle-point criteria $a(q)c(q)-b^2(q)$ because of cancellation, as we explicitly showed at Eq.~(\ref{AF_acb}).

\end{document}